\documentclass[10pt]{iopart}
\usepackage{iopams}
\usepackage{graphicx}
\usepackage[margin=1in]{geometry}
\include{jpeg2ps}

\begin{document}
\title{Tangled up in Spinning Cosmic Strings}
\author{Reinoud Jan Slagter}
\address{ASFYON, Astronomisch Fysisch Onderzoek Nederland, Bussum, The Netherlands \\ and The University of Amsterdam, The Netherlands  }
\ead{info@asfyon.com}
\begin{abstract}
It is known for a long time that the space time around a spinning cylindrical symmetric compact object such as the cosmic string, show un-physical behavior, i.e., they would possess closed time like curves (CTC). This controversy with Hawking's chronology protection conjecture is unpleasant but can be understood if one solves the coupled scalar-gauge field equations and the matching conditions at the core of the string. A new interior numerical solution is found of a self gravitating spinning cosmic string with a U(1) scalar gauge field and the matching on the exterior space time is revealed.
It is conjectured  that the experience of CTC's close to  the core of the string is exceedingly unlikely. It occurs when the causality breaking boundary, $r_\mu$, approaches the boundary of the cosmic string, $r_{CS}$. Then the  metric components become singular and the proper time on the core of the string stops flowing. Further, we expect that the angular momentum $J$ will decrease due to the emission of gravitational energy triggered by the scalar perturbations.
When a complete loop is taken around the string, the interior time jumps by a factor $2\pi J$. The proper time it takes to make a complete loop becomes infinite
and will be equal to the period that $g_{\varphi\varphi}$ remains positive. In this time interval  the angular momentum will be reduced to zero by emission of wave energy.
The physical situation of an observer who experience $r_{\mu}\rightarrow r_{CS}$  is very unpleasant: the energy-momentum tensor components diverge.
\end{abstract}
\maketitle
\section{Introduction}\label{Intro}	
In general relativity theory (GRT) one can construct solutions which are related to real physical objects. The most famous one is the black hole solution. One now believes
that in the center of many galaxies there is a rotating super-massive black hole, the Kerr black hole. Because there is an axis of rotation, the Kerr solution is a member of the family of the axially symmetric solutions of the Einstein equations\cite{Steph:2009}.
A legitimate question is if there are other axially or cylindrically symmetric asymptotically flat solutions of the equations of Einstein with a classical or non-classical matter distribution and with correct asymptotical behavior just as the Kerr solution. Many attempts are made, such as the Weyl-, Papapetrou- and Van Stockum solution\cite{Islam:1985}. None of these attempts result is physically acceptable solution.

It came as a big surprise that there exists a vortex-like solution in GRT comparable with the magnetic flux lines in type II superconductivity. These vortex lines occur as topological defects in an abelian U(1) gauge model, where the gauge field is coupled to a charged scalar field\cite{Niels:1973}.
It can  easily be established that the solution must be cylindrically symmetric, so independent of the z-coordinate and the energy per unit length along the z-axis is finite. The coordinate r now measures the distance from the z-axis. In fact, when one goes around a distant closed curve  in a non-cylindrically symmetric configuration  and we shrink the curve to a point, we produce a discontinuity in the phase factor, contradicting the smoothness of the Higgs field.
The static finite energy configuration cannot be stable, since we can press it down to the vacuum. So the jump is not allowed. In the cylindrically symmetric situation, the curve  cannot be shrunk to a point. Using Stoke's theorem, then the jump is again related to the magnetic flux in the string, i.e., $\frac{2\pi n}{e}$. One says that the abelian Higgs model is topological stable. This model, for a single flux quantum, is also known as the Nielsen-Olesen string\cite{Fels:1981}.
The U(1) vortex solution possesses  mass, so it will couple to gravity. When we incorporate the abelian U(1)gauge model into GRT, many of the features of the Nielsen-Olesen vortex solution and superconductivity will survive. However, there is also surprisingly new behavior of the resulting gravitating string.
There are two types, local (gauged) and global cosmic strings. We are mainly interested in local cosmic strings, because in a gauge model,  strings were formed during a local symmetry breaking and so have a sharp cutoff in energy, implying no long range interactions.
If we map the degenerated vacuum expectation values of the Higgs field to position space, one obtains now a locus of trapped energy in points through spacetime, i.e., a cosmic string\cite{Vil:1994}.
It is conjectured that in any field theory which admits cosmic string solutions, a network of strings inevitable forms at some point during the early universe. However, it is doubtful if they persist  to the present time. Evidence of these objects would give us information at very high energies in the early stages of the universe.
It turns out that in the early stages of the universe, when the temperature decreases, the scalar field develops spontaneous symmetry breaking. This results in the topological defects, as described above. The mass and dimension of the cosmic string is largely determined by the energy scale at which the phase transition occurred.
Cosmic strings could have served as seeds for the formation of galaxies. It is believed that the grand unification (GUT) energy scale is about $10^{16}GeV$. The mass per unit length of a cosmic sting will be of the order of $10^{18}$kg per cm, which is proportional to the square of the energy breaking scale. The length could be unbounded long, but its thickness is still a point of discussion. By treating the cosmic string as an infinite thin mass distribution, one will encounter serious problems in general relativity. This infinite thin string model give rise to the "scaling solution", i.e., a scale-invariant spectrum of density fluctuations, which in turn leads to a scale invariant  distribution of galaxies and clusters.
Cosmic strings can collide with each other and will intercommute to form loops. These loops will oscillate and loose energy via gravitational radiation and decay. There are already tight constraints  on the gravitational wave signatures  due to string loops via observations of the millisecond pulsar-timing data, the  cosmic  background radiation (CMB) by LISA and analysis of data of the LIGO-Virgo gravitational-wave detector. Its spectrum will depend on the string mass $G\mu$, where $\mu$ is the mass per unit length. Recent observations from the COBE, Wamp and Planck satellites put the value of $G\mu <10^{-7}$. It turns out that cosmic strings can not provide a satisfactory explanation for the magnitude of the initial density perturbations from which galaxies and clusters grew. The interest in cosmic strings faded away, mainly  because of the inconsistencies with the power spectrum of the CMB. Moreover, they will produce a very special pattern of lensing effect, not found yet by observations. The recently discovered "spooky" alignment of quasar polarization over a very large scale\cite{Huts:2014} good be well understood by the features of the cosmic strings and could be the first evidence of the existence of these strings.

New interest in cosmic strings arises when it was realized that cosmic strings could be produced within the framework of string theory inspired cosmological models. Physicists speculate that extra spatial dimensions could exist in addition to our ordinary 4-dimensional spacetime. These so-called cosmic super-strings can play the role of cosmic strings in the framework of string theory or M-theory, i.e., brane world models. Super-symmetric GUT's can even demand the existence of cosmic strings.
These theories can also be used to explain several of the shortcomings of the Standard Model, i.e., the unknown origin of dark energy and dark matter and the weakness of gravity (hierarchy problem). In these theories, the weakness of gravity might be fundamental. One might naively imagine that these extra dimensions must be very small, i.e., curled up and never observable.
Super-massive strings with $G\mu >1$, could be produced when the universe underwent phase transitions at energies much higher than the GUT scale.
Patterns of symmetry breaking can lead to monopoles, domain walls or cosmic strings.
Recently there is growing interest in the so-called brane world models,  first proposed by  Arkani-Hamed et. al.,\cite{ADD:1998},\cite{ADD1:1999} and extended by Randall and Sundrum (RS)\cite{RS:1999}. In these models, the extra dimension can be very large compared to the ones predicted in string theory. The difference with the standard super-string model is that the compactification rely on the curvature of the bulk.
We live in a (3+1)-dimensional brane, embedded in a 5-dimensional bulk spacetime. Gravitons can then propagate into the bulk, while the other fields are confined to the branes. The weakness of gravity can be understand by the fact that it "spreads" into the extra dimension and only a part is felt in 4D. This means that all of the four forces could have similar strengths and gravity only appears weaker as a result of this geometric dilution. The huge discrepancy between the electro-weak scale, $M_{EW}=10^3 GeV$ and the gravitational mass scale $M_{Pl}=10^{19} GeV$ will be suppressed by the volume of the extra dimension, or the curvature in that region. The effective 4D gravitational coupling will be $M_{eff}^2=M_{EW}^{n+2}R_0^n$, instead of a fundamental $M_{Pl}^2$. For $n=2$ and $M_{eff}$ of order $10^{19} GeV$, the compactification  radius $R_0$ will be of order of millimeters. This effect can also be achieved in the RS models by a warp factor $M_{eff}^2=(1-e^{-R_0})M_5^3$ on the visible brane at $y=L$, where y is the extra dimension. In the RS-2 model there are two branes, the visible and the gravity brane at y=0. The branes have equal and opposite tensions. The positive brane tension has fundamental scale $M_5$ and is hidden. If $M_5$ is of the order $R_0^{-1}\sim $TeV, we can recover $M_{Pl}\sim 10^{16}$TeV by choosing $\frac{L}{R_0}$ large enough\cite{shir:2000}.
At low energy, a negative bulk cosmological constant will prevent gravity to leak into the extra dimension, $\Lambda_5 = \frac{-6}{R_0^2}=-6\mu^2$, with $\mu$ the corresponding energy scale. The $\Lambda_5$ squeeze the gravitational field closer to the brane at $y=0$.
In the RS-1 model, one pushes the negative tension brane $L\rightarrow \infty$. If one fine-tunes the $\lambda =\frac{3M_{Pl}^2}{4\pi R_0^2}$, then this ensures a zero effective cosmological constant on the brane.  The infinite extra dimension makes, however, a finite contribution to the 5D volume due to the warp factor.
Because of the finite separation of the branes in the RS-2 model, one obtains so-called effective 4D modes ( KK-modes) of the perturbative 5D graviton on the 4D brane. These KK-modes will be massive from the brane viewpoint. In the RS-1 model the discrete spectrum disappears and will form a continuous spectrum.
Although the coupling of the KK-modes with matter will be very weak, it would be possible to find experimental evidence of these KK modes. With the new data from the Large Hadron collider at CERN, it might be possible to observe these extra dimensions and the electro-weakly coupled KK modes in the TeV range, at least for the RS-2 model. For the RS-1 model, the contribution of the massive KK-modes sums up to a correction of the 4D potential. For $r\ll L$ one obtains a slightly increase of the strength.
\begin{equation}
V(r)\approx\frac{GM}{r}\Bigl(1+\frac{2L^2}{3r^2}\Bigr).\label{Eq.1}
\end{equation}
For experiments on Newton's law, one find that $L\leq 0.1$ mm.
It will be clear that compact objects, such as black holes and cosmic strings, could have tremendous mass in the bulk, while their warped manifestations in the brane can be consistent with observations. So brane-world models could overcome the observational bounds one encounters in cosmic string models. $G\mu$ could be warped down to GUT scale, even if its value was at the Planck scale. Although static solutions of the U(1) gauge string on a warped spacetime show significant deviation from the classical solution in 4D\cite{slagter1:2012},\cite{slagter2:2012}, one is interested in the dynamical evolution of the effective brane equations. In the 4D case, it was found\cite{greg1:1997},\cite{greg1a:1999} that time dependency might remove the singular behavior of global cosmic strings.
One conjectures that, in contrast with earlier investigations\cite{sas1:1999}, that the wavelike back-reaction of the Weyl tensor on the brane and the quadratic corrections of the energy momentum tensor of the scalar-gauge field will have a significant impact on the evolution of the effective brane.
In the 4D case\cite{Greg1b:1988} investigations were done of  the effect of an infinite cosmic string on an expanding cosmological background. It turns out that the asymptotic spacetime is conical, just as the pure cosmic string spacetime. This is not desirable. Further, one can use the C-energy to estimate the cosmological gravitational radiation from these strings. It turns out that the produced cylindrical gravitational waves fade away during the expansion and becomes negligible. The corrections to the scaled solution appear at order $\frac{r_{cs}}{R_H}$, the ratio of the string radius to the Hubble radius. Clearly this ratio is extremely small ($10^{-20}$).
However, in a warped 5D setting, one can expect observable imprint of these so-called warped cosmic strings on the time evolution of the brane for values of the symmetry breaking scale much larger than the GUT values. The warp factor makes these strings consistent with the predicted mass per unit length on the brane. It seems that in these models there is a wavelike energy-momentum transfer to infinity on the brane induces via the disturbances of the cosmic string\cite{slagter4:2014}. Fluctuations of the brane when there is a U(1) scalar-gauge field present, are comparable with the proposed brane tension fluctuations ("branons"), whose relic abundance can be a dark matter candidate. It was found that on a warped FLRW background\cite{slagter5:2015} that brane fluctuations can be formed dynamically due to the modified energy-momentum tensor components of the U(1) scalar-gauge field. This effect is triggered by the time-dependent warp factor. The accelerated expansion of our universe could be explained without the cosmological constant, which would solve the controversial huge discrepancy between the the cosmological constant and the vacuum energy density.

Another unsolved problem in GRT is the possibility of formation of closed timelike curves (CTC).
At first glance, it seems possible to construct in GRT causality violating solutions. CTC's suggest the possibility of time-travel with its well-known paradoxes. Although most physicists believe that Hawking's chronology protection conjecture holds in our world, it can be tantalizing to investigate the mathematical underlying arguments of the formation of CTC's. There are several spacetimes that can produce CTC's. Most of them can easily characterized as un-physical. There is one spacetime, i.e., the Gott two cosmic string situation, which gained much attention the last several decades\cite{star:1963},\cite{deser1:1983},\cite{deser2:1992},\cite{gott:1990},\cite{hooft2:1992}. Two cosmic strings, approaching each other with high velocity, could produce CTC's. If an advanced civilization could manage to make a closed loop around this Gott pair, they will be returned to their own past.
However, the CTC’s will never arise spontaneously from regular initial conditions through the motion of spinless cosmons ( “Gott’s pair”):  there  are boundary conditions that has CTC’s also at infinity  or at an initial  configuration!
If it would be possible to fulfill  the CTC condition at $t_0$ then at sufficiently large times  the cosmons will have evolved so far apart that the CTC’s would disappear.
Gott’s CTC comes towards the interaction region of spacelike infinity, and so unphysical\cite{deser2:1992},\cite{hooft2:1992}.
Moreover, it turns out that the effective one-particle generator has a tachyonic centre of mass. This means that the energy-momentum vector is spacelike. Even a closed universe will not admit these CTC's\cite{hooft3:1993}. In fact, a configuration of point particles admits a Cauchy formulation within which no CTC's are generated.
So the Gott spacetime solution violates physical boundary conditions or the CTC's were preexisting. The chronology protection conjecture seems to be saved for the Gott spacetime.

There are still some unsatisfied aspects in these arguments. In order to study the Gott-spacetime, one usually omits in the metric the $dz^2$ term. The resulting conical (2+1)-dimensional spacetime is manifest locally flat outside the origin of the "source". But there is still the geometrical gravitational effect of the gravitating point particle located on  the place where the cosmic string intersects the $(\rho -\varphi )$-plane. When one transforms the effective spinning point-particle solution from the (2+1) dimensional spacetime to the (3+1) dimensional spacetime by adding the $dz^2$ term, one adds a Killing vector $\frac{\partial}{\partial z}$ to the spacetime.
One introduces in this (2+1) dimensional spacetime ad hoc an energy-momentum tensor. One must realize, however, that this specific solution depends crucial on the energy-momentum tensor of the scalar-gauge field in (3+1) dimensional spacetime. Some authors introduce the line singularity  by defining $T^{00}=\mu\delta^2(r)$. However, the U(1) cosmic string  has a thickness $\sim \frac{1}{\eta}$ and depends on the ratio of the masses of the scalar field and gauge field. So the delta function must be smeared out and the internal spacetime must be correctly matched onto the exterior spacetime.
In the construction of the Gott spacetime, the effective one-string generator obtains intrinsic angular momentum and its spacetime is of the Kerr-type( intrinsic spinning "cosmon")
\begin{equation}
ds^2=-(dt+J d\varphi )^2+dr^2+(1-4G\mu)^2 r^2d\varphi^2.\label{eq2}
\end{equation}
If one transforms this metric to Minkowski minus a wedge, then we have a helical structure of time: when $\varphi$ reaches $2\pi$, t jumps by $8\pi G J$. This metric has a singularity because
$T^{00}\sim 4G\mu\delta^2(r), T^{0i}\sim J\epsilon ^{ij}\partial_j\delta^2(r)$, describing a spinning point source. So in the extended (3+1) spacetime there is a CTC if for small enough region $J>(1-4G\mu)r_0$.
One can "hide" the presence of the spinning string by suitable coordinate transformation in order to get the right asymptotic behavior, but then one obtains a helical structure of time, not desirable.
For the Kerr solution, these CTC's are hidden behind the horizon, but not for this spinning cosmic string. So the spinning cosmic string solution possesses serious problems.

The Kerr  and spinning cosmic string spacetimes are both  members of the stationary axially symmetric spacetime. This spacetime can formally be obtained from the corresponding cylindrically symmetric spacetime by the complex substitution $t\rightarrow iz, z\rightarrow it , J\rightarrow iJ $. This spacetime admits gravitational waves. So could the CTC's be created by a dynamical process?
It is clear from the considerations in this introduction, that it seems worth  to investigate the spinning cosmic strings as a compact object in more detail when there is a U(1) scalar-gauge field is present. Further, one must match the interior  space time on an exterior space time with the correct junction conditions and correct asymptotic properties.
In section 2 we will give an overview of the stationary spinning string and in section 3 we introduce a new numerical solution.

\section{The stationary spinning string and the CTC dilemma}\label{sec2}
\subsection{Historical notes}\label{2:1}
Let us consider  the spacetimes of rigidly rotating axially symmetric objects\cite{Islam:1985}.
In polar coordinates one writes the axially symmetric spacetime as
\begin{equation}
ds^2=F(dt-Wd\varphi )^2-\frac{r^2}{F}d\varphi^2-e^\mu (dr^2 +dz^2),\label{eq3}
\end{equation}
with $F, W$ and $\mu$ functions of  r and z.
The Papapetrou\cite{papa1} solution is
\begin{eqnarray}
F=\frac{1}{\alpha\cosh \Bigl(\frac{z}{(z^2+r^2)^{3/2}}\Bigr)-\beta\sinh \Bigl(\frac{z}{(z^2+r^2)^{3/2}}\Bigr)}\qquad W=-\frac{\sqrt{\alpha^2-\beta^2}r^2}{(z^2+r^2)^{3/2}},\label{eq4}
\end{eqnarray}
where $\mu$ is obtained by quadratures. This solution is asymptotically flat, because $W$ has already the correct asymptotic form of that of a rotating body  and
\begin{equation}
F\rightarrow \frac{1}{\alpha} +\frac{\beta z}{\alpha^2r^3}-\frac{3}{2}\frac{\beta z^3}{\alpha^2 r^5}+.....\label{eq5}
\end{equation}
for large r. We see that $F$ has the correct asymptotically form, but there is no term proportional to $\frac{1}{r}$. So there is no mass term.
If there is no rotation, $W=0$, we obtain the Weyl solution\cite{weyl}.
\begin{equation}
F=\Bigl(\frac{z-z_0+\sqrt{r^2+(z-z_0)^2}}{z+z_0+\sqrt{r^2+(z+z_0)^2}}\Bigr)^\mu,\label{eq6}
\end{equation}
It represents the gravitational potential of a thin uniform rod with density $\mu$ and z-dimensions $(-z_0,z_0)$. It has the correct asymptotic form and a mass term
\begin{equation}
F=1-\frac{2\mu m}{r}+\frac{2\mu^2 m^2}{r^2}+....\label{eq7}
\end{equation}
This can be seen by calculating the gravitational potential, which becomes $\mu\delta(x)\delta(y)$, with $\delta $ the Dirac delta function.
For $\mu =1$ we obtain the Schwarzschild solution. Now we should like to combine the Weyl solution ( non-rotating) and the Papapetrou solution in order to get the right asymptotic form. This is hard to manage, due to the fact that rotating sources produce extra gravitational effects.

Close related to this problem is the infinitely long rigidly rotating cylindrical symmetric dust solution of Lewis-van Stockum\cite{stockum}. When we rewrite the metric in the form
\begin{equation}
ds^2=Fdt^2-H(dr^2+dz^2)-Ld\varphi^2-2Md\varphi dt,\label{eq8}
\end{equation}
then the exterior solution becomes ( independent of z)
\begin{eqnarray}
H=e^{-a^2R^2}(\frac{R}{r})^{2a^2R^2},\quad L=\frac{1}{2}r R sinh(3\epsilon +\theta) csch(2\epsilon) sech(\epsilon),\label{eq9}
\end{eqnarray}
\begin{eqnarray}
M=r sinh(\epsilon +\theta) csch(2\epsilon),\quad   F=\frac{r}{R} sinh(\epsilon -\theta) csch(\epsilon),\label{10}
\end{eqnarray}
with $\theta =\sqrt{1-4a^2R^2}\ln(\frac{r}{R})$ and $ tanh(\epsilon)=\sqrt{1-4a^2R^2}$. The Lewis-van Stockum solution has asymptotically not the correct behavior, so it cannot represent the exterior field of a bounded rotating source. Nevertheless it can be used by generating new solutions.    The mass and angular momentum per unit z-coordinate are $\mu=\frac{1}{2}a^2R^2$ and
$J=\frac{1}{4}a^3R^4$ respectively.
It is remarkable that only for $aR<\frac{1}{2}$ this metric can be transformed to a local static form by the transformation $t\rightarrow a_1t+a_2\varphi, \quad \varphi\rightarrow a_3\varphi +a_4 t$.
But then time coordinate becomes periodic. So this metric has manifestly CTC's.
The resulting metric is that of Levi-Civita
\begin{equation}
ds^2=r^{2C}dt^2-A^2r^{2C^2-2C}(dr^2+dz^2)-r^{2-2C}d\varphi^2.\label{eq11}
\end{equation}
It contains two constants, whereas the Newtonian solution contains only one. C is related to the mass per unit length, or angle deficit as in the case of a cosmic string. The two constants are fixed by the internal composition of the cylinder, just as in the case of the cosmic string where the solution is determined by the symmetry breaking scale and the gauge-to-scalar mass ratio.
The dependance of the exterior solution on two parameters has a strong bearing on the existence of gravitational waves. After the complex transformation to the cylindrical symmetric metric, $t\rightarrow iz, z\rightarrow it, W\rightarrow iW$ one obtains the two formally equivalent forms
\begin{eqnarray}
ds^2=F(dt-Wd\varphi )^2-\frac{A^2}{F}d\varphi^2-e^\mu (dr^2+dz^2)\cr
ds^2=-F(dz-Wd\varphi)^2-\frac{A^2}{F}d\varphi^2-e^\mu(dr^2-dt^2).\label{eq12}
\end{eqnarray}
For example, the counterpart solution of the static axially symmetric Weyl solution, is the Einstein-Rosen wave-solution. In general, the presence of aperiodic gravitational waves should be attributed to radiation of gravitational radiation from the core of the mass. One parameter, related to the angular momentum, should return to its original value, when the system underwent a symmetric motion for a limited  period of time, while the inertial mass parameter will decrease. This is also the case for a cosmic string: the angle deficit will decrease after emission of gravitational energy\cite{slagter1:2001}. A lot of investigation was done by Belinsky and Zakharov\cite{Bel1:1978},\cite{Bel2:1979}, to find a generalization of the Einstein-Rosen wave solution in the general case of two dynamical degrees of freedom, in stead of one as in the case of the Einstein-Rosen solution. One can also use the concept of the Ernst potential and the analogies existing between gravitational waves and solutions with cylindrical symmetry, to find solutions for the rotating cosmic string which is radiating\cite{Xan1:1986},\cite{Xan2:1986},\cite{Eco:1988}.
They are of the Petrov type D and represent soliton-like gravitational waves interacting with a cosmic string. The solution are obtained by using as "seed" the Minkowski metric.
In the latter case, it turns out that the  metric tends slower to asymptotic flatness when one approaches infinity along a null direction than a spacelike one. For spacelike infinity $r>>|t|$, asymptotic flatness can only be achieved  by a change in the z-coordinate of the form $z\rightarrow z+pd\varphi$ in order to obtain orthogonality of the killing fields $\frac{\partial}{\partial\varphi}$ and $\frac{\partial}{\partial z}$, which is equivalent after the complex substitution to a periodic time. One cannot switch off rotation: it  will give rice to global effects.
Near the symmetry axis, $r<<|t|$ ( not near the light cone $|t|=r$), the metric approaches a conical spacetime. The C-energy\cite{Thorne:1965} is constant and there is no energy flux.
Far from the symmetry axis, $r>>|t|$, the angle deficit differs from the one obtained near $r=0$ and there is also no energy flux and the C-energy is again constant but smaller than the one near the symmetry axis.
For $r=|t|\rightarrow\infty$ there is incoming radiation from past null infinity and is equal the outgoing radiation at future null infinity. This indicates that there is a gravitational disturbance propagating along the null cone $|t|=r$.
Although it is clear that intervening gravitational waves do contribute to the angle deficit. For $r\rightarrow 0$ the metric component $\frac{g_{\varphi\varphi}}{r^2}$ don't approach a constant value. So again,  the concept of a rotating string fades away in this approach. One cannot maintain the classification of a rotating string as an expression of the effective action of the radiating gravitational field at spacelike infinity plus a static cosmic string at the axis $r=0$.

\subsection{The stationary spinning cosmic string}\label{2:2}
Let us consider the stationary cylindrical symmetric spacetime with angular momentum $J$
\begin{equation}
ds^2=-e^A\Bigr[(dt+Jd\varphi)^2-dz^2\Bigr]+dr^2+K^2e^{-2A}d\varphi^2,\label{eq13}
\end{equation}
with $A, K$ and $J$  functions of r.

The starting point is the Lagrangian\cite{lag1:1987}, \cite{lag2:1989}, \cite{garf1:1985}
\begin{equation}
{\cal L}=\frac{1}{16\pi G}R-\frac{1}{2}D_\mu\Phi(D^\mu\Phi)^*-V(\mid\Phi\mid)-\frac{1}{4}F_{\mu\nu}F^{\mu\nu},\label{eq14}
\end{equation}
where $D_\mu\equiv\nabla_\mu+ieA_\mu$, $F_{\mu\nu}=\nabla_\mu A_\nu -\nabla_\nu A_\mu$, $ V(\Phi)=\frac{\beta}{8}(\mid\Phi\mid^2-\eta^2)^2$ and  $\eta$ the energy scale of symmetry breaking.
For GUT-scales, $\eta\sim 10^{16}GeV$, leading to a thickness of $\delta = \eta^{-1} \sim 10^{-30} cm$. The scalar and gauge fields take the form
\begin{eqnarray}
\Phi=Q(r,t)e^{in\varphi}\qquad A_\mu=\frac{n[P(r,t)-1]}{e}\nabla_\mu\varphi,\label{eq15}
\end{eqnarray}
with n the winding number the scalar field phase warps around the string. Further, one has for the masses $ m_\Phi ^2=\beta \eta^2$ and $ m_A^2=e^2\eta^2$, so $ \frac{m_A^2}{m_\Phi^2}=\frac{e^2}{\beta}\equiv\alpha$.
The parameter $\alpha$ is also called the Bogomol'nyi parameter. If $\alpha$ is taken 1, the vortex solution is super-symmetrizable.
If one re-scales $Q\equiv\eta X,  r\rightarrow\frac{r}{\eta\sqrt{\beta}}$ and $ K\rightarrow\frac{K}{\eta\sqrt{\beta}}$, then the radii of the core false vacuum and magnetic field tube are $r_\Phi\approx 1, r_A^2\approx\frac{1}{\alpha}$ and one has only two free parameters $\alpha$ and
$\eta$.
There don't exist a solution in closed form.
From the scalar-gauge field equations on the metric of Eq.(13) we then obtain
$\partial_rJ\partial_rP=0$ and $J$=constant$=J_0$,
so the field equations become identical to the classical diagonal case. So the metric can be transformed to Minkowski minus a wedge if we make the identification $t\rightarrow t+J\varphi$. So there is the 'helical' structure of time. Otherwise the metric suffers boost invariance along the symmetry axis.

One can easily proof\cite{garf1:1985} that in the case $J_0=0$ the metric outside the core is
\begin{equation}
ds^2=-e^{a_0}(dt^2-dz^2)+dr^2+e^{-2a_0}(k_2r+a_2)^2d\varphi^2,\label{eq16}
\end{equation}
where $a_0$ and $a_2$ are integration constants.
So the metric field $K$ can be approximated for large $r$ by $(k_2r+a_2)$, where $k_2$ will be determined by  the energy scale $\eta$ and
the ratio $\frac{m_A}{m_\Phi}=\sqrt{\alpha}$.
This metric can be brought to Minkowski form by the change of variables
\begin{eqnarray}
r'=r+\frac{a_2}{k_2}, \varphi '=e^{-a_0}k_2\varphi, t'=e^{a_0/2}t, z'=e^{a_0/2}z,\label{eq17}
\end{eqnarray}
where now $\varphi '$ takes values $0\leq\varphi'\leq 2\pi e^{-a_0}k_2$. So we have a Minkowski metric minus a wedge with angle deficit
\begin{equation}
\Delta\theta=2\pi(1-e^{-a_0}k_2).\label{eq18}
\end{equation}
It is obvious that the mass per unit length and so the angle deficit, depends on the behavior of the scalar and gauge fields.
When one increases the symmetry breaking scale above the GUT scale $G\mu\approx 10^{-6}$, then the properties of the stringlike solution change drastically. Beyond a certain value the gravitational field becomes so strong that it restores the initial symmetry and the string could become singular at finite distance of the core. These super-massive cosmic strings, predicted by superstring theory, are studied because the universe may have undergone phase transitions at scales much higher than the GUT scale, i.e., $G\mu\geq 1$. The angle deficit increases with the energy scale, so when it becomes greater than  $2\pi$, the conical picture disappears and is replaced by a Kasner-like metric. There are some exceptions. In the case of the Bogomol'nyi bound, i.e., $8\beta =e^2$, one can find a regular solution for large values of $\mu$, representing a vortex where the spacetime is the product of ${\bf R}^2$ by a 2-surface which is asymptotically a cylinder\cite{linet1:1990}. In general, it will be difficult to establish gravitational Bogomol'nyi inequalities due to the non-local nature of gravitational energy. For the abelian Higgs global string case it turns out that a solution exists for the Bogomol'nyi inequality\cite{comtet:1987}.
Numerical analysis of super-massive cosmic strings\cite{lag2:1989}, where $G\mu\gg 10^{-6}$, shows that the solution becomes singular at finite distance of the core  of the string, or the angle deficit becomes greater than $2\pi$.
These features could also arise if the coupling between the scalar and gauge field is very weak\cite{ortiz:1991}. These low-energy super-massive strings are closely related to the U(1) global strings.
On  warped 5D spacetime  this picture changes significantly\cite{slagter6:2012,slagter5:2015}

In our spinning case ($J_0\neq 0$), the angle deficit now depends on $J_0$
\begin{equation}
\Delta\theta =2\pi\Bigl[1-\lim_{r\rightarrow\infty}\frac{d}{dr}(\sqrt{K^2e^{-2A}-e^A J_0^2})\Bigr].\label{eq19}
\end{equation}
In figure 3 we plotted the component $g_{\varphi\varphi}$ for $J_0=0$ ( static case) and $J_0\neq 0$ ( stationary case).
\begin{figure}[h]
\centerline{
\includegraphics[width=5.5cm]{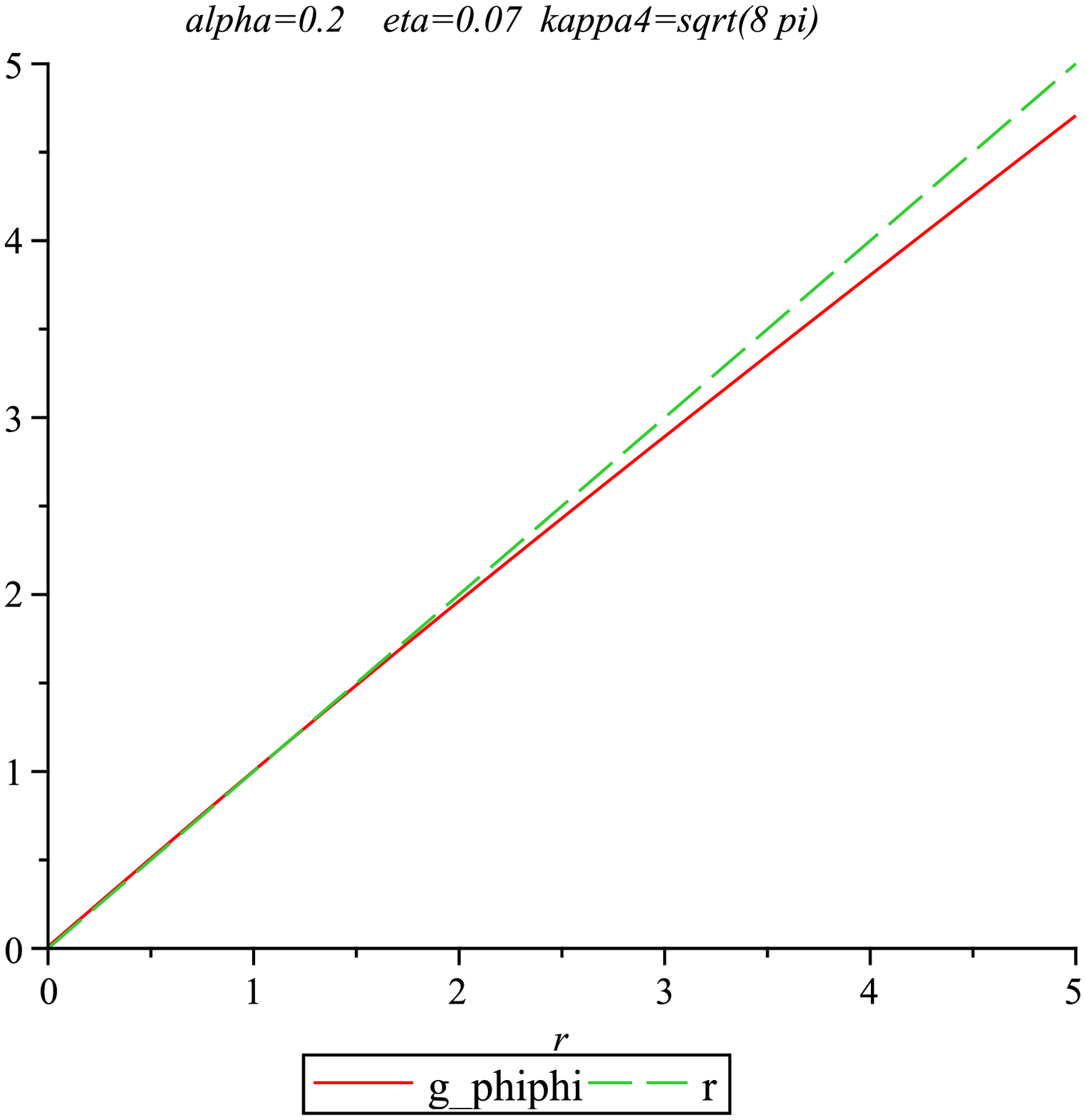}
\includegraphics[width=5.5cm]{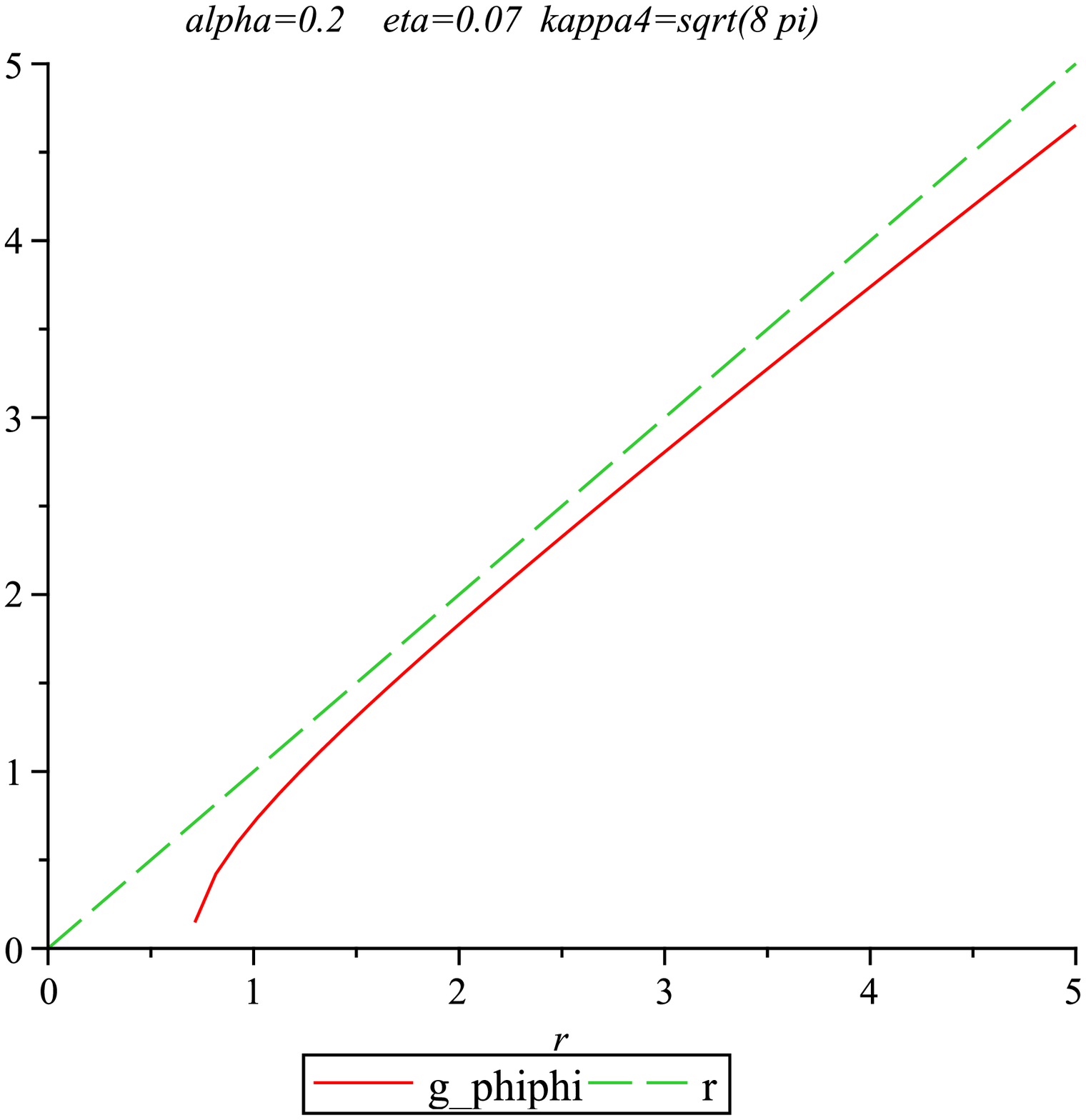}}
\vskip 1.cm
 \caption{The angle deficit for $J_0=0$ (left) and $J_0\neq 0$}
\label{fig:3}
\end{figure}
We see that there is a lower bound ( for $K\approx r$):  $r> e^{1.5A}J_0$. So can the mass of the spinning string be confined within radius $r_0$, such that $J_0<e^{-1.5A}r_0$ in order to avoid CTC's?
For large r there is no significant change in behavior.
It is remarkable that for the U(1) gauge string the angular momentum must be r-independent. So the correct asymptotic behavior cannot be accomplished without the peculiar time-helical structure.
The question remains how to go in the axially symmetric situation from the spinning cosmic string to the static one without the introduction of the periodic time and with the emission of gravitational waves.
\subsection{ The boundary problems at the core of the cosmic string}\label{sec2:3}
What will happen with the asymptotic structure of a time-dependent solution of the cosmic string when one starts with a CTC-free initial situation?
Already mentioned in section (2.2), a smooth distribution of matter is often replaced by a concentrated source, such as a cosmic string, with no internal structure. In the case of electromagnetism there is a mathematical framework for this idealization, because the Maxwell field and the charge-current density can be treated as distributions by virtue of their linearity. Such a framework cannot be applied to Einstein's equations, being non-linear. It is hard to introduce in GR gravitating point particles, i.e., sources concentrated on a one  dimensional surface in spacetime. Thin shells of matter, concentrated on a three-dimensional surface can be constructed in GR for suitable boundary and continuity conditions across the boundary\cite{Israel:1966}. However, the stress-energy of the sheet acts as a source, so there will be gravitational radiation due to the jump in the first derivative of the metric crossing the boundary. The Weyl tensor concentrated on this surface yields the amplitude of the radiation\cite{Chan:1984}. What happens if we consider sources concentrated on a two-dimensional surface in spacetime? Behave  these strings more like shells or point particles? An illustrative example of the problems one encounters is the Gott-Hiscock exact interior and exterior solution of the U(1) gauge string\cite{Gott:1985},\cite{Hiscock:1985}. This metric is given by
\begin{equation}
ds^2=dt^2+dz^2+dr^2+b(r)^2d\varphi^2,\label{eq20}
\end{equation}
with $b(r)$ given by
\begin{eqnarray}
b(r)=a\sin(\frac{ r}{a}),\quad  (r<r_0)  \qquad b(r)=\cos\frac{r_0}{a}r , \quad (r>r_0).\label{eq21}
\end{eqnarray}
For $r>r_0$ we have the usual Minkowski spacetime with angle deficit and for $r<r_0$ we have non-zero components of the energy-momentum tensor $T_z^z =T_t^t =\frac{1}{8\pi G a^2}$. This is comparable with our cosmic string solution of section (2.1).
If one demands continuity at the boundary, then $a\rightarrow \infty $\cite{Ray2:1990}, which makes the energy stress tensor vanish throughout the string. If one allows a discontinuity, then $T_\mu^\nu$ blows up at the boundary. It turns out that this solution represents a homogeneous stationary universe with with a magnetic field and cosmological term,comparable with a Melvin universe.
If we take $r_0\rightarrow 0$, we obtain a line source with line mass density $\mu$ equals the angle deficit and with $T^{tt}\sim \mu\delta (x)\delta (y)$. So one should conclude that we have found a solution of a line source. But one can easily  proof\cite{Geroch:1987} that the line distribution of mass in the Newtonian limit results in a  zero external field. So one needs some detailed restrictions on the distribution of the matter content of a cosmic string in order to overcome these serious problems. This could be done by imposing gravitational radiation or to keep $r_0\neq 0$. It was found\cite{Garf2:1990} that strongly gravitating zero-thickness vacuum cosmic strings can be described, as long as the stress-energy remains bounded as the string is approached,  by a pp-traveling-wave solutions  of Einstein's equations. If the stress-energy is given by the scalar-gauge field of the U(1) model, then it is impossible to apply this approach: one has to fulfil also the scalar-gauge field equations from a zero thickness singular line source.

If one considers a global string, where during the phase transition the global symmetry is broken, it was found that the spacetime has a curvature singular, not removable due to a bad choice of coordinates\cite{Greg1b:1988,Greg1c:1996}. The main reason for this fact is that the one of the energy momentum tensor components becomes singular. In general, the global string has a less well-defined boundary than the local string.
By considering the time-dependent extension, the singularity could possibly be removed\cite{greg1:1997}. Moreover, it is questionable if the global string has asymptotically a conical spacetime\cite{Ray2:1990}.
As mentioned before, the spinning string must have a boundary separating the interior vortex solution from the exterior conical spacetime. Before we consider the time-dependent situation, some note must be made about the behavior of the boundary layer in the case of metric Eq.(13) if we introduce a boundary at $r=r_s$  (for A=0) and without the scalar-gauge field as matter content. When  one approaches the boundary from the interior one encounters a serious problem in smoothly matching $J(r)$ to a constant $J$, if $\partial_r J \neq 0$\cite{janca:2007}.
Moreover, without the specific mass of the scalar-gauge field, one can prove in this case that the weak energy condition is violated for an observer at $r=r_s$ for suitable four-velocity. Some additional fields must be added to compensate for the energy failure close to $r_s$, which can be  the U(1) scalar-gauge field.
However, for the scalar-gauge field equations we shall see that a numerical solution can be obtained with physical acceptable boundary behavior.
\section{A New Numerical Solution of the Spinning Cosmic String}\label{sec7}
Let us now consider a different approach to this problem. Consider again the interior metric:
\begin{equation}
ds^2=-e^{A}\Bigl[(dt+Jd\varphi )^2-dz^2\Bigr]+dr^2+\frac{K^2}{e^{2A}}d\varphi^2,\label{eq22}
\end{equation}
where A, K and J are now functions of r and t.
Further, we consider the scalar and gauge fields $R$ and $P$ also time-dependent. From the scalar-gauge field equations one obtains $ \partial_t(RP)=0$. For $R=0$, we are dealing with a "Melvin-type" spacetime. Let us consider here $P=0$ (global string).   We then obtain  from the scalar equations the spin-mass relation
\begin{equation}
J(r,t)=Z K(r,t)e^{-2A(r,t)},\label{eq23}
\end{equation}
with Z a constant and the partial differential equations
\begin{equation}
R_{tt}=-\frac{R_tK_t}{H}+\frac{e^{2A}}{(e^A-Z^2)}\Bigl\{R_{rr}+e^{-2A}(e^A-2Z^2)R_tA_t +\frac{1}{K}R_rK_r+\frac{1}{2}\lambda R(\eta^2-R^2)\Bigr\},\label{eq24}
\end{equation}
\begin{eqnarray}
K_{tt}&=&\frac{1}{(e^A-Z^2)}\Bigl\{(e^A-3Z^2)K_tA_t+2e^{2A}A_rK_r+\frac{3}{2}K(Z^2A_t^2-e^{2A}A_r^2)\cr &+&2\pi GKe^{2A}\Bigl(\lambda(R^2-\eta^2)^2+4(e^{-2A}Z^2R_t^2-R_r^2)\Bigr)\Bigr\},\label{eq25}
\end{eqnarray}
\begin{eqnarray}
A_{tt}=\frac{1}{2(e^A-2Z^2)}\Bigl\{(4e^A-5Z^2)A_t^2-e^A(4Z^2+3e^A)A_r^2+\frac{4e^A(Z^2+e^A)}{K}A_rK_r\cr +\frac{2(Z^2-2e^A)}{K}A_tK_t -\frac{Z^2e^A}{K^2}K_r^2
+4\pi G\Bigl(\lambda e^{2A}(R^2-\eta^2)^2+4(Z^2+e^A)R_t^2-4e^{2A}R_r^2\Bigr)\Bigr\}.\label{eq26}
\end{eqnarray}
These equations are consistent with $\nabla_\mu T^{\mu\nu}=0$. The equations for $K$ and $A$ don't contain the second order derivatives with respect to r. For the numerical solution we  used a slightly different set of PDE's.
\begin{figure}[h]
\centerline{
\includegraphics[width=5.5cm]{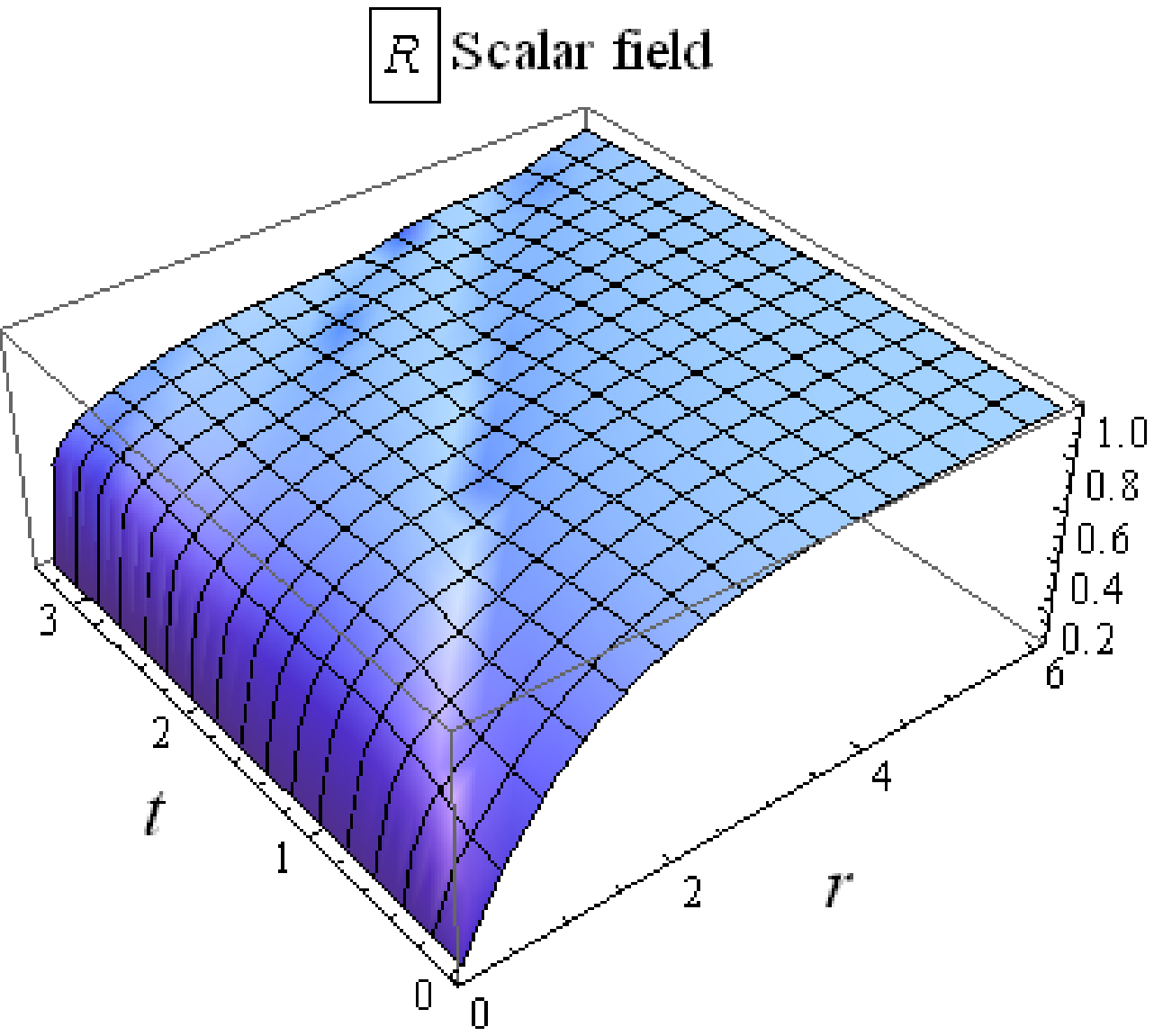}
\includegraphics[width=5.5cm]{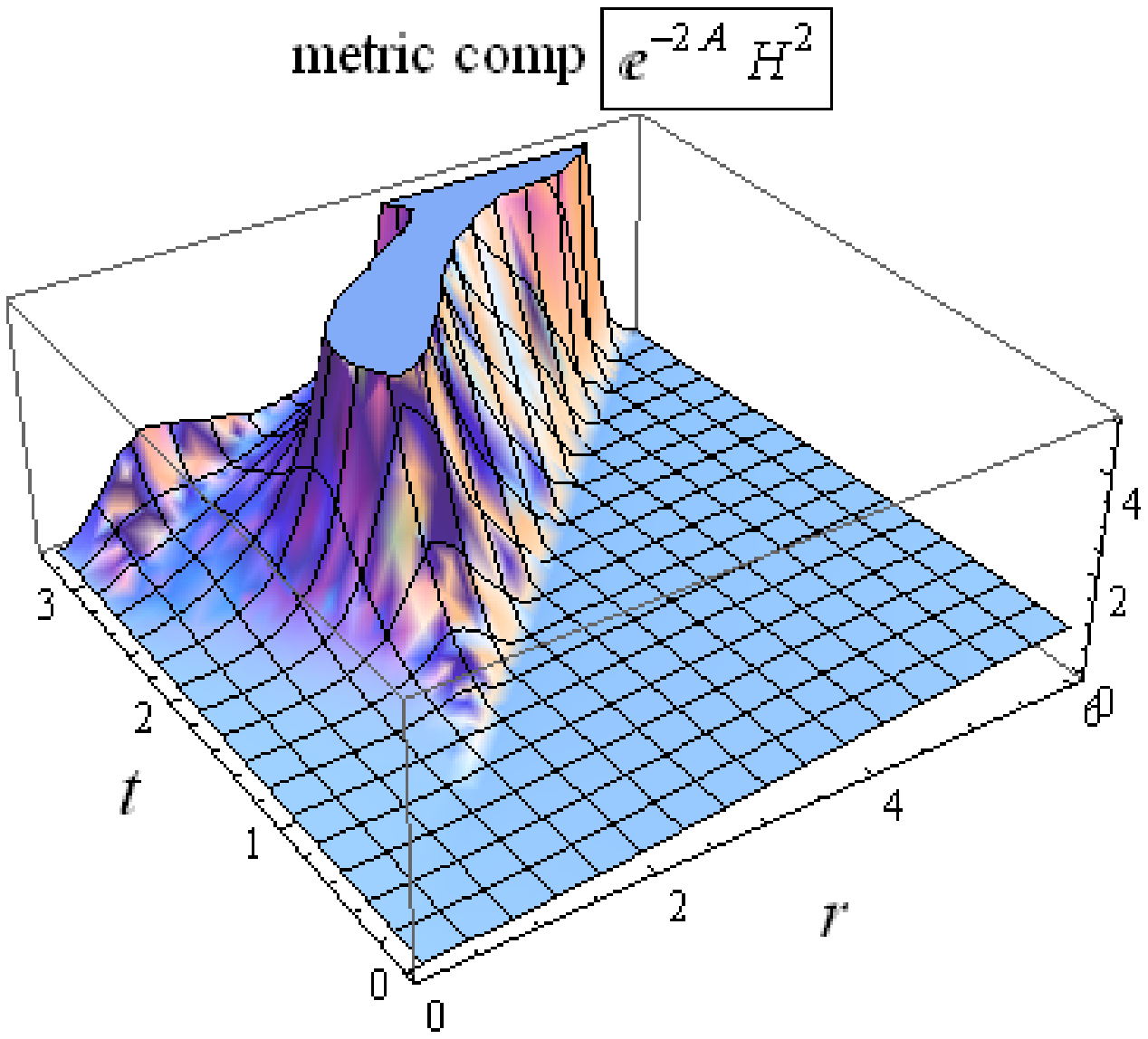}}
\centerline{
\includegraphics[width=5.5cm]{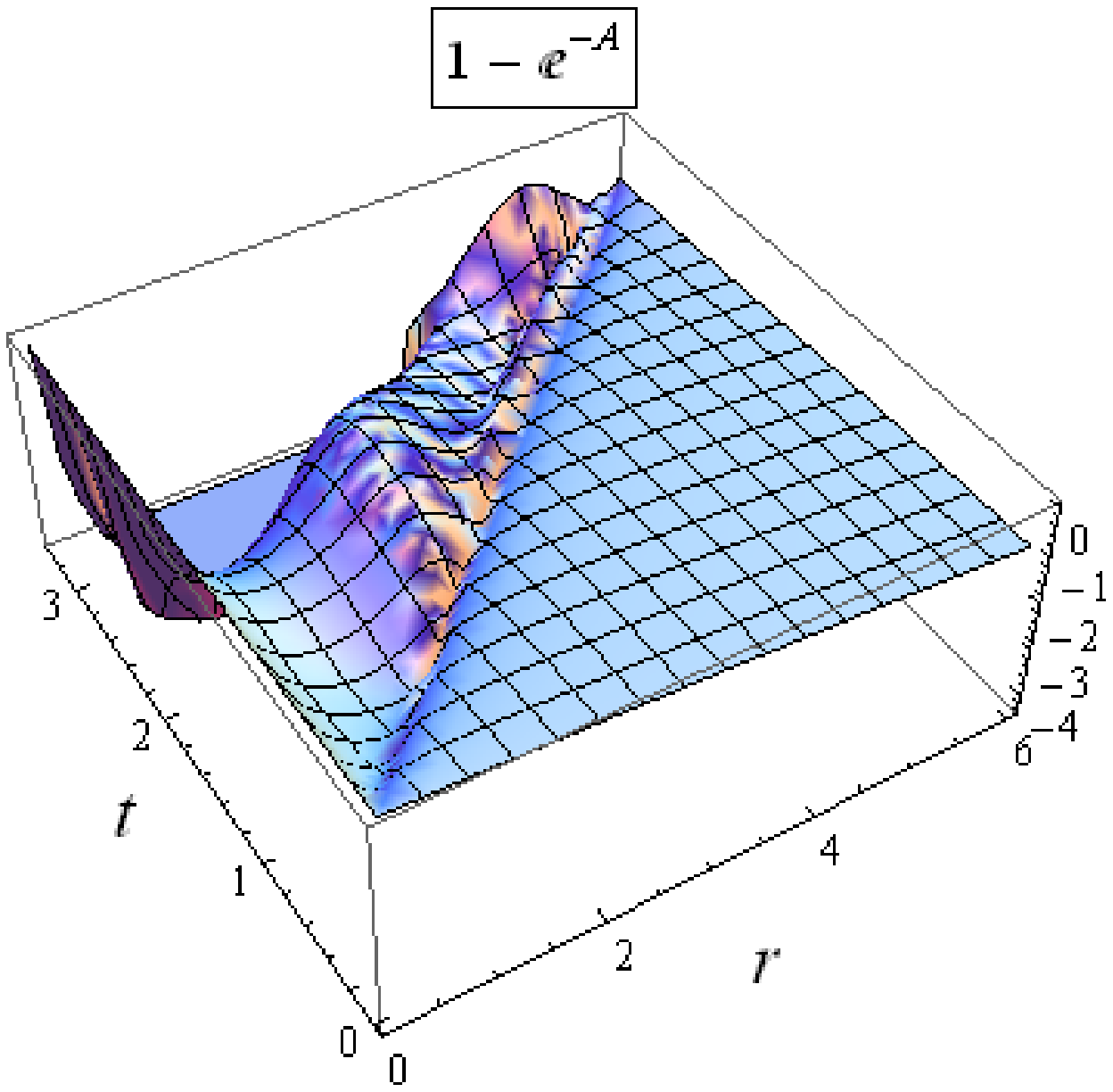}
\includegraphics[width=5.5cm]{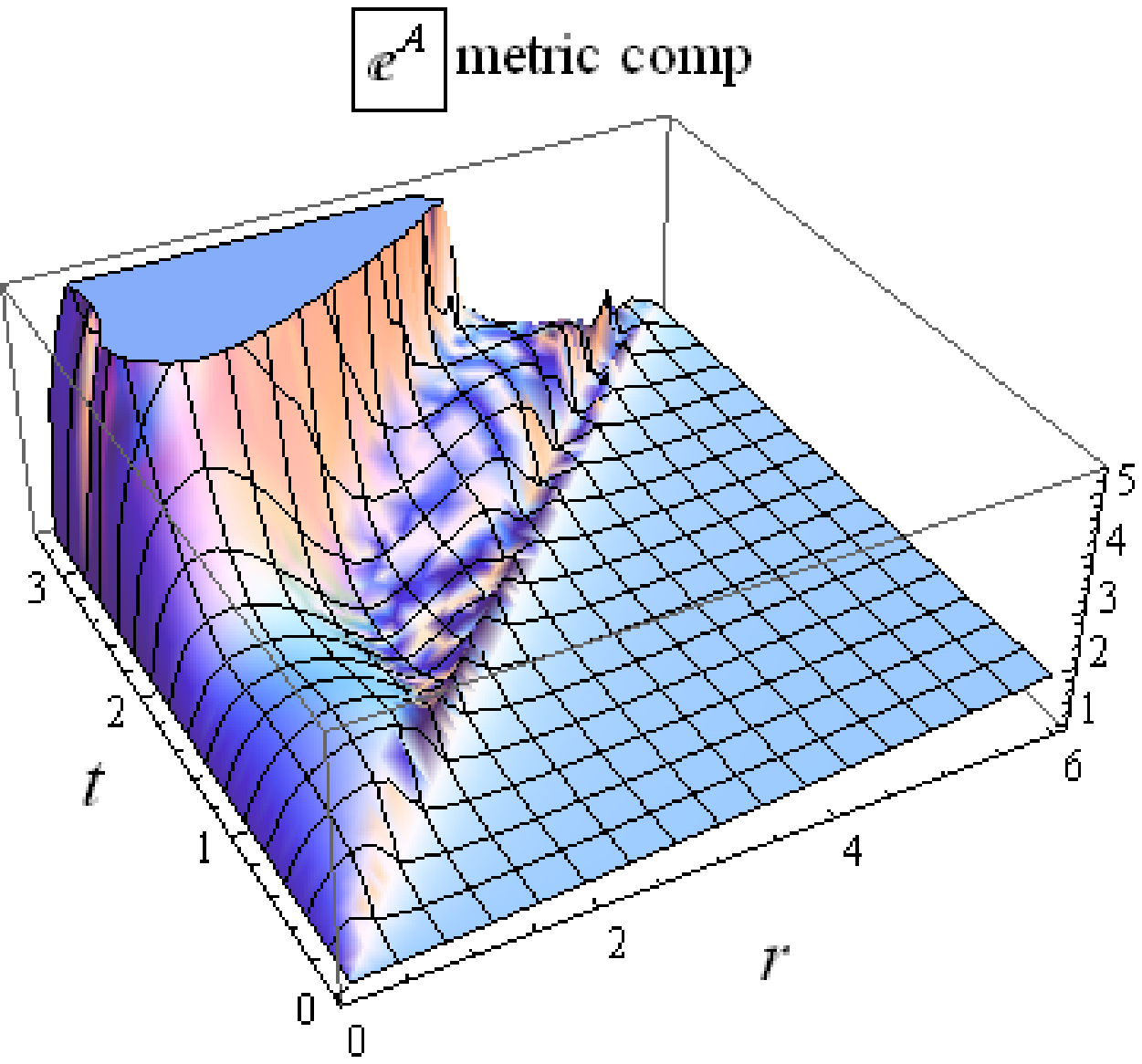}}
\centerline{
\includegraphics[width=5.5cm]{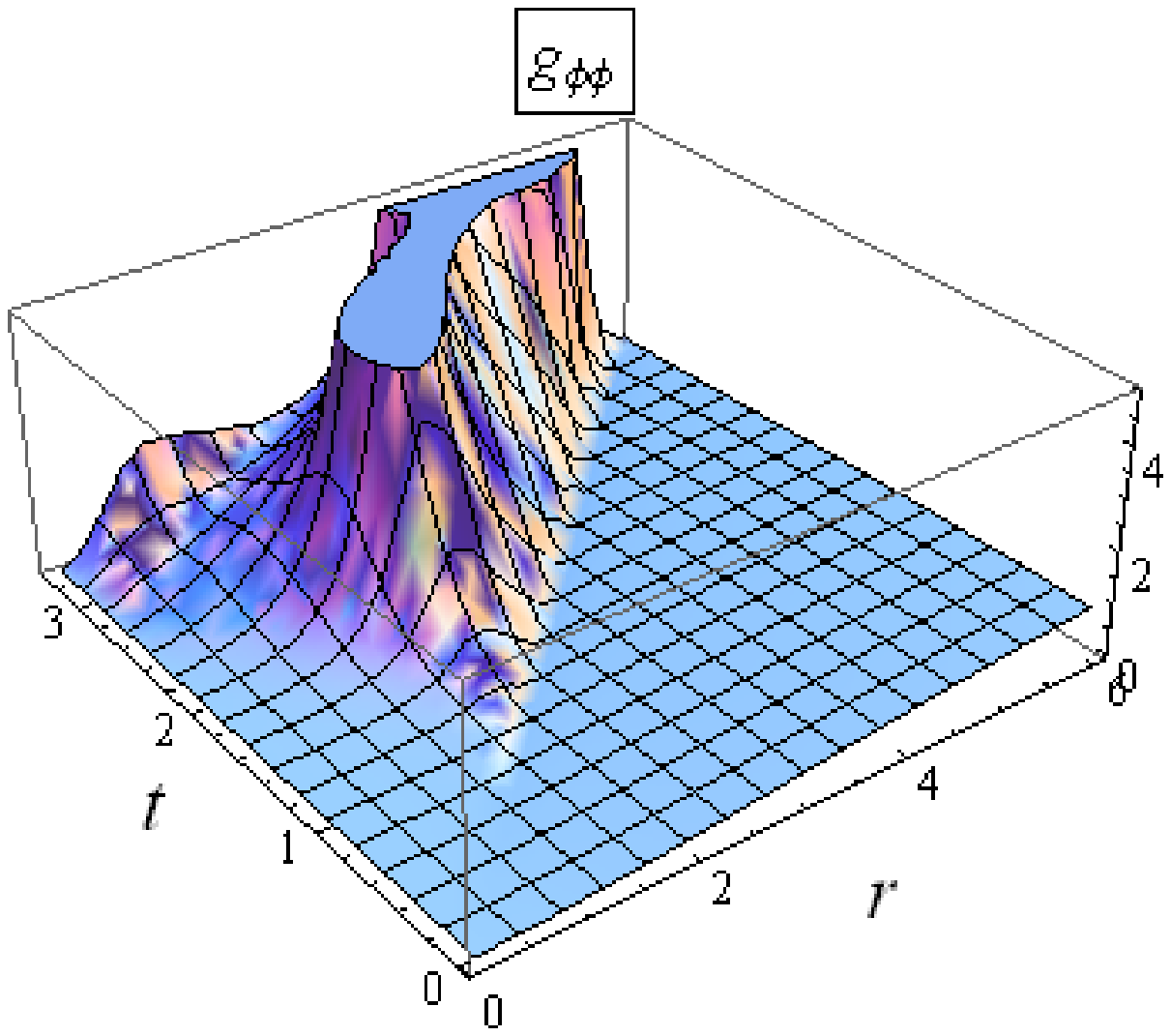}
\includegraphics[width=5.5cm]{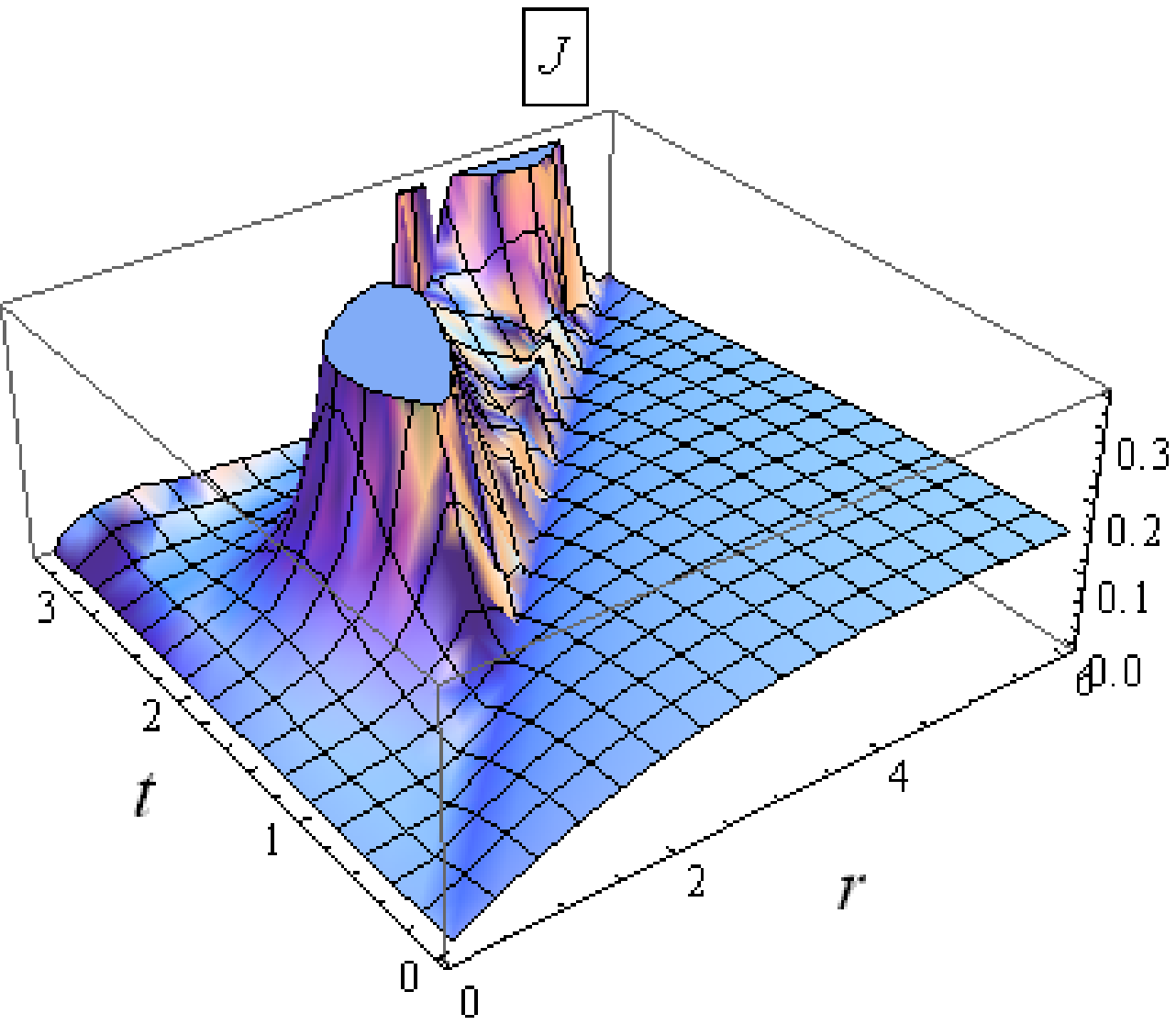}}
\centerline{
\includegraphics[width=5.5cm]{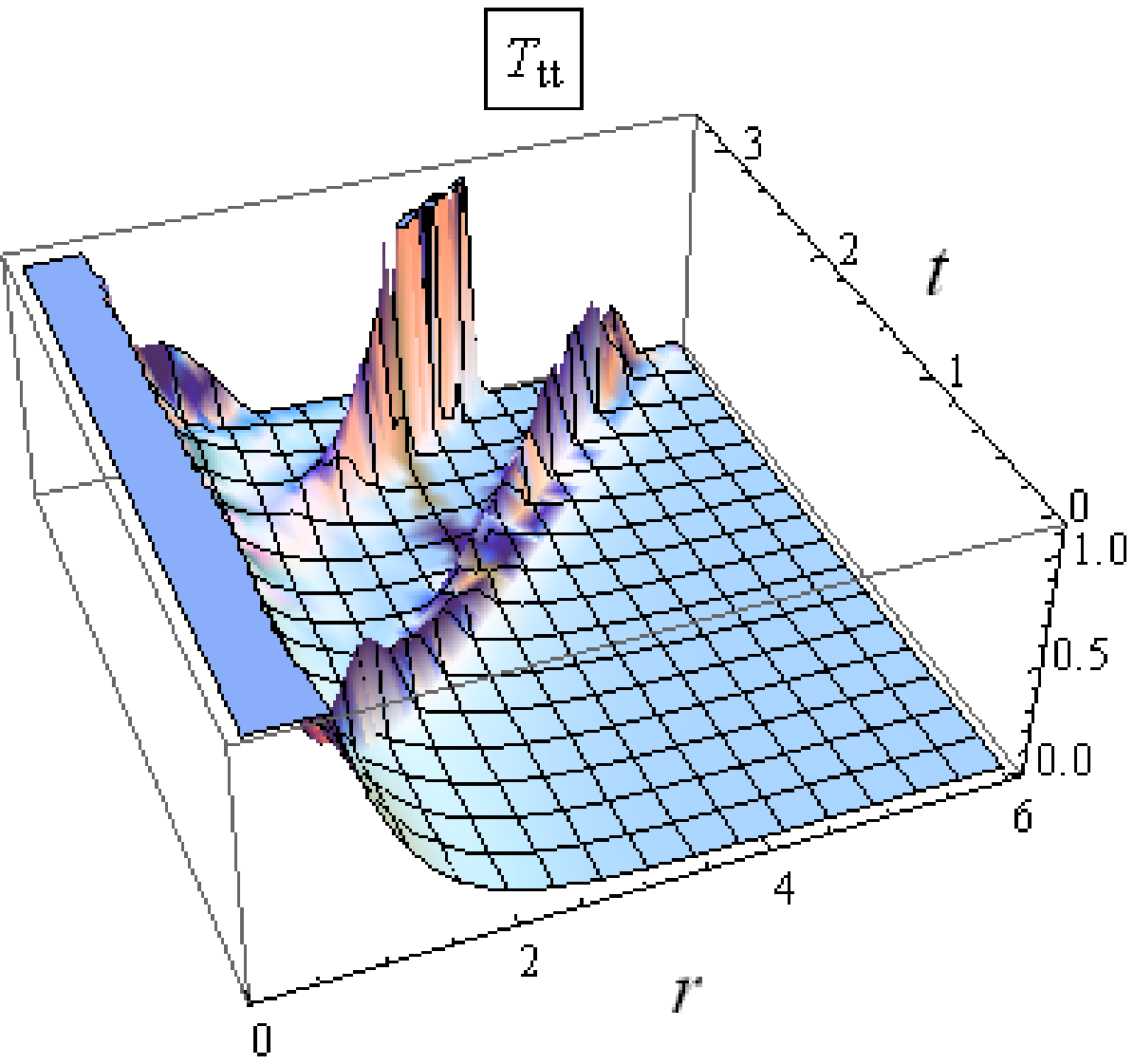}
\includegraphics[width=5.5cm]{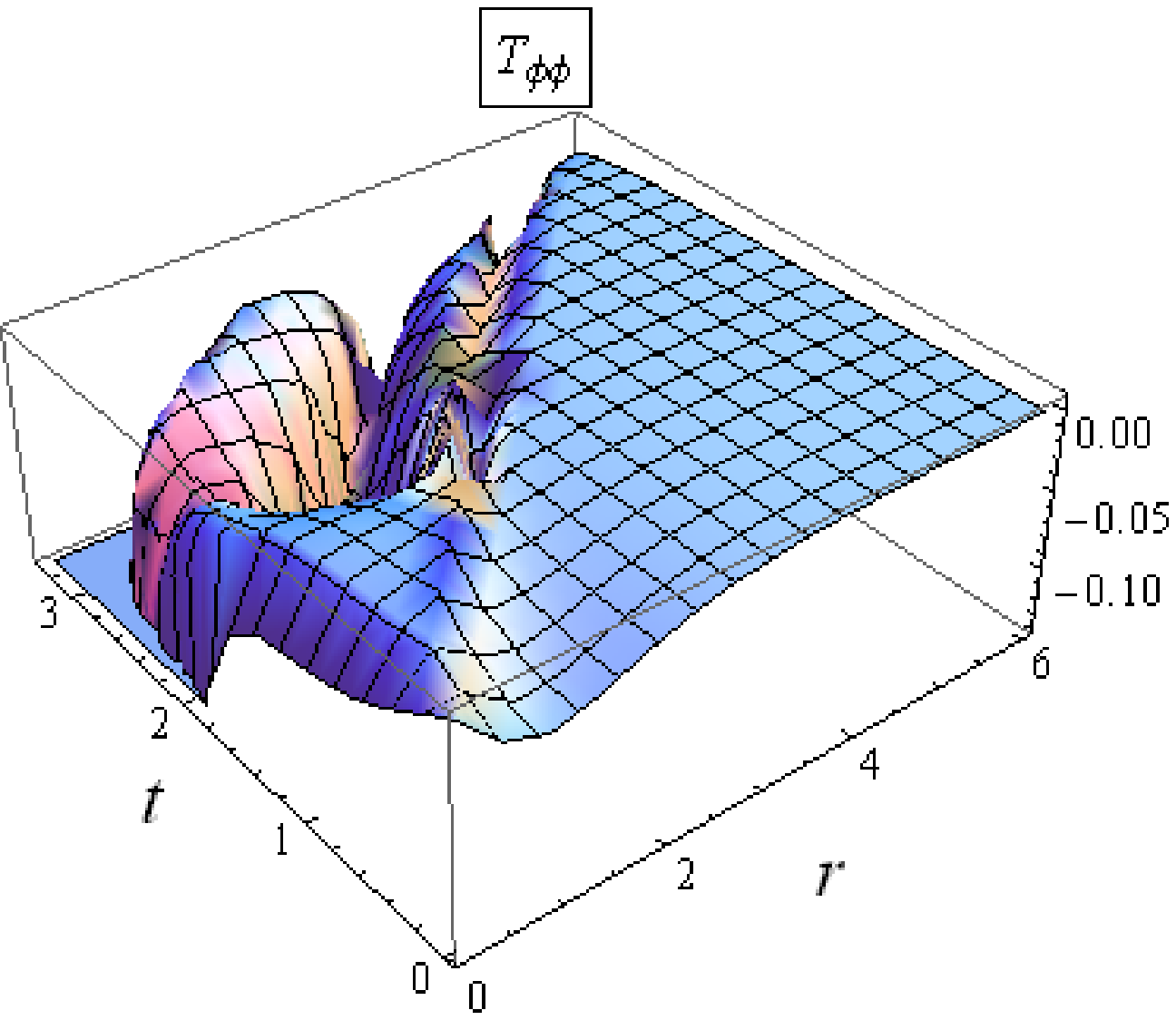}}
 \caption{Typical time-dependent solution for the interior spacetime with angular momentum $J=ZKe^{-2A}$.}
\label{fig:4}
\end{figure}
Is is easily verified from the scalar-gauge field equations  that in the case of t-independency of the scalar and gauge field, $J_{t}=0$, so the case of section (2.2) is obtained. This contradicts earlier results\cite{soleng1:1992}.

Further, the curvature scalar $R$, can change sign when $J<Ke^{-1,5A}$, so is not strictly positive as in the stationary non-spinning global string case. This condition is just the non-CTC criterium of Eq.(22).
In Figure 2 we plotted a typical solution, where we choose for the initial $J(0)$ a 'kink'-solution tanh(r). We observe a singular boundary moving outward. This is the causality breaking boundary, separating the regions of $g_{\varphi\varphi}<0$ and $>0$. This singular behavior is also observed in the field equations Eq.(24), Eq.(25) and Eq.(26): for $e^A=Z^2$ or $e^A=2Z^2$.

For the exterior metric we have
\begin{equation}
ds^2=-e^{A}\Bigl[(dt+J_0d\varphi )^2-dz^2\Bigr]+dr^2+B(r)^2d\varphi^2.\label{eq27}
\end{equation}
For $A=0$ one can match the two solutions at the core of the string at $r_s$.
As discussed in section (2.2), the exterior metric must be written in diagonal form by the transformation $t=t^*-J_0\varphi$, with $t^*$ the exterior time and t the interior time, i.e.,
\begin{equation}
ds^2=-dt^{*2}+dz^2+dr^2+B(r)^2d\varphi^2.\label{eq28}
\end{equation}
For  GUT strings we have $B(r)\rightarrow (1-4G\mu)(r)$, with $\mu$ the energy density of the string. The metric can further be transformed to the flat conical spacetime ( see section (2)) with angle deficit $8\pi G\mu$. The radius $r_s$ of the core of the string must satisfy $r_s>\frac{J_0}{1-4G\mu}$ as long as $\mu <\frac{1}{4G}$, otherwise will $g_{\varphi\varphi} <0$.
From Eq.(23) we then obtain the matching condition at $r_s$ ($e^A=1$)
\begin{equation}
Z=\frac{J_0}{(1-4G\mu)r_s},\label{eq29}
\end{equation}
so $Z<1$. We see that for $Z^2=e^A$ and $Z^2=\frac{1}{2}e^A$ the equations for K, A and R become singular. So if $Z<1$ then also $e^A $ can become smaller than 1.  This means that the proper time of a test particle becomes smaller than the coordinate time. The situation $e^A <1$ occurs in the local string situation when the parameter $\alpha =m_{gauge}/m_{scalar}$, i.e., the ratio of the masses of the gauge field and the scalar field, becomes also $<1$. This completes our understanding of the transition of a local string to a global string.
Let us define now $r_\mu =\frac{J_0}{(1-4G\mu)}$. Then $Z=\frac{r_\mu}{r_{CS}}$. So when  $Z<1$, then $r_\mu < r_{CS}$ and the CTC resides inside the core of the string when we may consider  $r_\mu$ as the the causality breaking boundary.

It was found in the case of the U(1) gauge string\cite{slagter:1996} that the metric component $g_{\varphi\varphi}$ becomes negative for suitable values of the parameters of the local string model, such as the VEV $\eta$ and $\lambda$. The smaller $\eta$, the later the negative region is encountered and the CTC resides inside the core of the string. In our global string situation we encounter two singular boundaries, as can be seen from the plot of $T_{tt}$ in figure 2.
Let us consider the hypersurface $\Sigma$, the boundary of the interior and exterior spacetime:
\begin{equation}
ds_\Sigma^2 = -\epsilon d\tau^2 +dz^2+r_{CS}^2(\tau)d\varphi^2,\label{eq30}
\end{equation}
with $\tau$ the proper time on the boundary. By applying the boundary conditions\cite{slagter:1996}, one finds evolution equations for $ r_{CS}$, and the shell's stress tensor, which can be expressed in the extrinsic curvature tensor  (Lanczos tensor). These equations can be solved numerically together with the relations $t=t^*-J\varphi$ and $\dot t^*=\sqrt{1+\dot r_{CS}^2}$. One can plot $r_{CS}$, for suitable values for the parameters of the model, as function of the interior time and proper time and compare it with the evolution of $r_\mu \equiv \frac{J}{1-4 G\mu}$. See figure 3.
\begin{figure}[h]
\centerline{
\includegraphics[width=6cm]{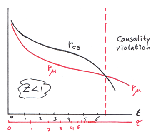}
\includegraphics[width=6cm]{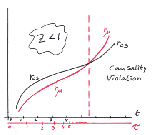}}
 \caption{Advanced (left)  and retarded evolution of the string core radius $r_s$ compared with $r_\mu =\frac{J}{1-4G\mu}$}
\label{fig:5}
\end{figure}

It is conjectured  that the formation of CTC's outside the core of the string is exceedingly unlikely. It occurs when $Z\rightarrow 1$, i.e., $r_{CS}\rightarrow r_\mu$. Then K and R become singular and the propertime
on the core of the string stops flowing. Further, we expect that $J_0$ will decrease due to the emission of gravitational energy triggered by the scalar perturbations.

When a complete loop is taken around the string ( so $\varphi$ acquires a phase shift of $e^{2in\pi}$), the interior time jumps by a factor $2\pi J_0$. The proper time it takes to make a complete loop becomes infinite
and will be equal to the period that $g_{\varphi\varphi}$ remains positive. In this time the angular momentum will be reduced to zero by emission of wave energy.

This is a different situation compared, for example, with  the Kerr solution, where the CTC is hidden behind the horizon. In the string case there is no horizon and to experience the CTC, one will encounter  on the core of the string a freezing of proper time and after a complete loop, the angular momentum will be reduced to zero. The physical situation of an observer who experience $r_{CS}\rightarrow r_\mu$  is very violent: the energy-momentum tensor components diverge.

\section{Conclusions}\label{sec8}
From a numerical solution of the field equations of the scalar- and metric fields, it is concluded that an observer travelling close to the boundary of a spinning cosmic string, will never experience a CTC. The proper time it takes to make a complete loop will become infinite. In the corresponding coordinate time, the increasing causality violating interior region meets the core radius of the string and a singular behavior is encountered for the metric components. The energy-momentum tensor components diverge, not a pleasant location to be part of.

\end{document}